\documentclass[a4paper,10pt]{article}

\usepackage{multirow}
\usepackage{booktabs}
\usepackage[dvips]{graphicx}
\usepackage{amsthm}
\usepackage{amssymb}
\usepackage{eepic}
\usepackage{amscd}
\usepackage{longtable}
\usepackage{array}

\newcommand{\ue}[2]{\mathop{\vphantom=#2}
\limits^{\scriptscriptstyle{#1}}}
\newcommand{\Spec}{{\rm Spec}}
\newcommand{\Z}{{\mathbb Z}}

\newcommand{\C}{{\mathbb C}}
\newcommand{\frs}{{\mathfrak s}}
\newcommand{\frt}{{\mathfrak t}}
\newcommand{\eps}{\epsilon}
\newcommand{\del}{\delta}
\newcommand{\one}{\mbox{1 \hspace{-3.2mm} {\bf \rm l}} }
\newcommand{\ot}{\otimes}
\newcommand{\brabra}{{\langle\!\langle}}
\newcommand{\ketket}{{\rangle\!\rangle}}
\newcommand{\Tr}{{\rm Tr}}

\begin{document}

\title{Remarks on
the multi-species exclusion process
with reflective boundaries}
\author{Chikashi Arita
\thanks{chikashi.arita@cea.fr}  }
\date{\ }

\maketitle

\begin{abstract}
We investigate one of the simplest 
multi-species generalizations 
of the one dimensional exclusion process
with reflective boundaries.
The Markov matrix governing the dynamics
of the system splits into blocks (sectors) specified by 
the number of particles of each kind.
We find matrices connecting the blocks 
in a matrix product form.
The procedure (generalized matrix ansatz)
 to verify that a matrix intertwines 
 blocks of the Markov matrix
was introduced in the periodic boundary condition,
which starts with a local relation
(Arita et al,
{\it J. Phys. A} \textbf{44}, 335004 (2011)).
The solution to this  relation 
for the reflective boundary condition
is much simpler than that for
the periodic boundary condition.
\end{abstract}

\section{Introduction}\label{Sec:Intro}

The asymmetric simple exclusion process (ASEP)
 is a lattice-gas model of interacting particles \cite{L},
where each particle is a random walker hopping from
  a site  to one of the neighboring  locations 
  only if the target site is empty.
The ASEP in one dimensional lattice
($\Z$ or its subset) has been intensively studied as 
an exact solvable non-equilibrium model \cite{BE,D,S2}
which is related to growth phenomena \cite{J,KPZ,SS} 
and applied to modeling of various transport systems
``from molecules to vehicles''
 \cite{MGP,SCN}.
One of standard generalizations of the ASEP
to multi-species systems
($N$-species ASEP)
 where each site $i$ 
takes a local state 
$k_i \in \{1, \ldots, N+1\}\, (N \ge 0)$
 is as follows;
 nearest neighbor pairs of local states 
$(\alpha, \beta)=(k_i, k_{i+1})$
are interchanged  
\begin{eqnarray}\label{rule}
\alpha \beta \to \beta \alpha
\quad \mbox{with rate}\quad
  \left\{ \begin{array}{ll}
    1 &  \ (\alpha < \beta),\\
    q &  \ (\alpha > \beta),
  \end{array}\right.
\end{eqnarray}
where we impose $0\le q\le 1$ without loss of generality.
We say that
 the site $i$ is occupied
   by an $\alpha${\it th-class particle} for $k_i=\alpha\le N$.
We also say that the site $i$ is empty
  for $k_i=N+1$, which is usually denoted by 0.
The usual ASEP corresponds to $N=1$.
We will be formally concerned with 
the zero-species ASEP ($N=0$) as well.
We will call the cases $q=1$ and $q=0$ 
 multi-species symmetric simple exclusion process (SSEP)
and multi-species totally ASEP (TASEP), respectively.

In \cite{AKSS}, the spectral structures 
of  Markov matrices (which govern the dynamics 
of the system) were clarified.
On the other hand, in \cite{PEM},
the matrix product form for the stationary state
was found in the periodic-boundary case.
Then these two studies were combined in \cite{AAMP1}
where matrices connecting dynamics
of different values of $N$ were constructed.
This generalizes the procedure of the matrix
(product) ansatz for stationary states
\cite{BE,DEHP,HPS}.
In particular this generalized matrix ansatz
 enables us to {\it transfer} information
  to a system consisting of $N$-species particles
 from simpler systems consisting
 of $N'(<N)$-species particles.
The question is whether this generalized matrix ansatz
 is applicable to the {\it reflective-boundary} case or not.
This will be answered positively,
which is the main purpose of this paper.

Here we define the model on the $L$-site chain precisely.
Let $\left\{ |1\rangle, \ldots,  |N+1\rangle \right\}$
   be the basis of the 
single-site space $\C^{N+1}$,  
and represent a particle configuration 
$k_1\cdots k_L$ as the ket vector
$ |k_1 \cdots k_L\rangle
 = |k_1\rangle \ot \cdots \ot |k_L\rangle
\in (\C^{N+1})^{\otimes L}$.
We also use the corresponding bra vectors  
$   \langle k| $ and 
$   \langle k_1 \cdots k_L|  
= \langle k_1| \ot \cdots \ot \langle k_L| $
with $ \langle  j | k \rangle =\del_{jk}  $.
In terms of the probability vector
\begin{eqnarray}
 |P(t)\rangle = \sum_{1\le k_i \le N+1}
 P(k_1\cdots k_L; t)|k_1\cdots k_L\rangle,
\end{eqnarray}
with each coefficient $P(k_1\cdots k_L; t)$
representing the probability of finding
the configuration $k_1\cdots k_L$ at time $t$,
our model is governed by the master equation
$\frac{d}{dt}|P(t)\rangle = M^{(N)} |P(t)\rangle$.
The matrix $M^{(N)}$ has the form
\begin{eqnarray}
\label{eq:Markov}
M^{(N)} &\!\!\!=& \!\!\! 
  \sum_{1\le i\le L-1} \left(M^{(N)}_{\rm Loc\ } \right)_{i,i+1}, \\
M^{(N)}_{\rm Loc} &\!\!\! =&\!\!\! 
  \sum_{\alpha,\beta=1}^{N+1}
   \left(-\Theta(\alpha-\beta)|\alpha\beta\rangle\langle\alpha\beta|
      +\Theta(\alpha-\beta)|\beta\alpha\rangle\langle\alpha\beta|\, \right),
\label{eq:MLoc}
\end{eqnarray}
where $ \left(M^{(N)}_{\rm Loc\ } \right)_{i,i+1}$ acts 
  nontrivially on the $i$th and $(i\!+\!1)$st 
  components of the tensor product,
and $\Theta$ corresponds to the transition rate
\begin{eqnarray}\label{rate}
\Theta(\alpha-\beta) = \left\{
\begin{array}{cc}
1& (\alpha < \beta),\\
q& (\alpha > \beta),  \\
0& (\alpha = \beta).
\end{array}\right. 
\end{eqnarray}
We call the $(N+1)^L\times (N+1)^L$ matrix
$M^{(N)}$ total Markov matrix
 or simply Markov matrix,
 and the $(N+1) ^2 \times (N+1)^2$
  matrix $M^{(N)}_{\rm Loc}$
 local Markov matrix.
The relevant two-dimensional vertex model
is a special case of Perk-Schultz model
  \cite{ADHR,AR,PS,Sc}. 

We emphasize that we will investigate the multi-species ASEP
on the $L$-site closed segment with
 the {\it reflective boundary condition}
({\it free boundary  condition}).
I.e., we do not impose any boundary term
in the Markov matrix (\ref{eq:Markov}).
The total Markov matrix obviously preserves 
the number of particles of each class
and thus it has the following block diagonal structure:
\begin{eqnarray}
\label{=otMm}
M^{(N)}=
 \bigoplus_{m}M_m, \quad M_m \in {\rm End} V_m,\quad
(\C^{N+1})^{\otimes L}=\bigoplus_{m}V_m.
\label{=otVm}
\end{eqnarray}
Here each sector $V_m$
is specified by the number of particles of each class
\begin{eqnarray}\label{Vm=span}
V_m=\bigoplus_{\# \{i|k_i=j \} = m_j}
\C |k_1\cdots k_L \rangle,\quad
\end{eqnarray}
for $m=(m_1,\dots,m_{N+1})$.
This means that each sector is spanned by
the ket vectors corresponding to permutations
of the sequence $ \underbrace{1\cdots 1}_{m_1}
\underbrace{2\cdots 2}_{m_2}\cdots
\underbrace{N+1\cdots N+1}_{m_{N+1}}$.
We also call the label $m$
 for the sector $V_m$  ``sector''.
We write the dual space for each sector as
\begin{equation}
   V^*_m=\bigoplus_{\# \{i|k_i=j \} = m_j}
\C \langle k_1\cdots k_L | .   
\end{equation}

This paper is organized as follows.
In section \ref{sec:symmetry} 
we review the symmetry of the model,
and construct matrices which connect 
dynamics of different sectors.
We  call these connecting matrices
 ``conjugation matrix'', and
in section \ref{sec:matrix-product}
we will 
 reconstruct them in a matrix product form.
In section \ref{sec:diffN} we consider
  a relation between Markov matrices
  of different values of $N$,
  by introducing similar conjugation matrices.
In section \ref{sec:stationary-state} 
we investigate the stationary state and 
the relaxation to it.
We will see that
the stationary state can be regarded
as a product of conjugation matrices.
Section \ref{sec:summary} is the summary of this paper.

\section{Symmetry}\label{sec:symmetry}

The Markov matrix is mapped to
a $U_q(SU(N+1))$
 invariant quantum Hamiltonian $H$
by the similarity transformation
 \cite{ADHR,AR,CP,Da,Sa,S1}
\begin{eqnarray}
 \label{eq:similarity}
 -  \frac{1}{\sqrt{q} }  {S} M^{(N)} {S}^{-1}
   =   H^{(N)}  ,\ 
  {S} = \sum_{1\le k_i\le N+1} 
  q^{\frac{1}{4}\sum_{1\le i<j\le L} {\rm sign} (k_i-k_j) }
  |k_1\cdots k_L \rangle\langle  k_1\cdots k_L|, \\
 H^{(N)} = \sum_{1\le i\le L-1} \left(H^{(N)}_{\rm Loc}\right)_{i,i+1}, \ 
H^{(N)}_{\rm Loc}
 = -\sum_{1\le \alpha,\beta\le N+1 \atop  \alpha\neq\beta}
   |\alpha\beta\rangle\langle\beta\alpha|
 + \sum_{1\le \alpha,\beta\le N+1}
  \frac{\Theta(\alpha-\beta)}{\sqrt{q}}
   |\alpha\beta\rangle\langle\alpha\beta| .
\label{eq:H}
\end{eqnarray}
Thus $M^{(N)} $ also has a symmetry,
which we review in this section.
We define $U_q (SU(N+1))$ generators 
\begin{eqnarray}
 f^{(n)} = | n+1 \rangle  \langle n |, \quad
e^{(n)} = | n \rangle  \langle n+1 |, \quad
k^{(n)}=
\sqrt{q} | n \rangle  \langle n |  +
\frac{1}{\sqrt{q}}   | n+1 \rangle  \langle n+1 |  
   + \sum_{1\le x\le N+1 \atop x\neq n,n+1}
    | x \rangle  \langle x | 
\end{eqnarray}
and the comultiplications
\begin{equation}
  \Delta f^{(n)} = f^{(n)}\ot Id + k^{(n)}\ot f^{(n)},\quad
 \Delta e^{(n)} = e^{(n)}\ot (k^{(n)})^{-1} + Id\ot e^{(n)},\quad
 \Delta k^{(n)} =  k^{(n)}\ot   k^{(n)},
\end{equation}
which commutes with 
the local Hamiltonian:
\begin{equation}\label{eq:[D,H]=0s}
 [  \Delta f^{(n)} , H^{(N)}_{\rm Loc}  ]  = 
 [  \Delta e^{(n)} , H^{(N)}_{\rm Loc}  ]  = 
 [  \Delta k^{(n)} , H^{(N)}_{\rm Loc}  ]  = 0.
\end{equation}
From these local relations, we find 
 global commutation relations
 \begin{equation}
 [  F^{(n)} , H^{(N)} ]  = 
 [  E^{(n)} , H^{(N)} ]  = 
 [  K^{(n)} , H^{(N)} ]  = 0 ,
\end{equation}
where
\begin{eqnarray}
  F^{(n)}  &\!\!\!\!=& \!\!\!\! 
  f_1^{(n)} + k_1^{(n)} f_2^{(n)} + \cdots
   + k_1^{(n)}\cdots k_{L-1}^{(n)} f^{(n)}_L, \\
  E^{(n)}  &\!\!\!\!=& \!\!\!\! 
 e_1^{(n)}(k_2^{(n)})^{-1} \cdots (k_L^{(n)})^{-1}+
 \cdots + e_{L-1}^{(n)}(k_L^{(n)})^{-1}+ e_L^{(n)}, \\
  K^{(n)} &\!\!\!\!=& \!\!\!\!
   k^{(n)}_1 \cdots k^{(n)}_L ,
\end{eqnarray}
and each $x_i^{(n)}\ (x=f,e,k)$
acts nontrivially on site $i$.
From the similarity transformation we have 
\begin{eqnarray}
\label{eq:[FM]=[EM]=[KM]=0}
 [ \widetilde  F^{(n)} , M^{(N)} ]  = 
 [ \widetilde  E^{(n)} , M^{(N)} ]  = 
 [ \widetilde  K^{(n)} , M^{(N)} ]  = 0,
\end{eqnarray}
where
 $\widetilde X^{(n)} = S^{-1} X^{(n)}S\ (X=F,E,K)$.
By the direct calculation,
one can show that
 the elements of $\widetilde F^{(n)}$
 and  $\widetilde E^{(n)}$ are  given as
\begin{eqnarray}\label{eq:F-element}
  \langle j_1\cdots j_L | \widetilde F^{(n)} 
  | k_1 \cdots k_L \rangle
  &\!\!\!\!=& \!\!\!\! \left\{\begin{array}{ll}
    Q q^{   \# \{ i'<i | j_{i'} = n  \}  }
    &   \begin{array}{l} \!\!\!(
 \exists i:  j_i-1=k_i=n, \\
   j_\iota = k_\iota (\iota\neq i)  ), 
\end{array}   \\
    0   & (\rm otherwise), 
  \end{array}\right. \\
  \langle k_1\cdots k_L | \widetilde E^{(n)} 
  | j_1 \cdots j_L \rangle
 &\!\!\!\!=& \!\!\!\!
  \left\{\begin{array}{ll}
  Q q^{   \# \{ i'>i  | k_{i'} = n+1  \}  }
    &   \begin{array}{l}\!\!\!(
    \exists i :j_i-1=k_i=n, \\
      j_\iota = k_\iota (\iota\neq i)  ),
  \end{array}   \\
    0   & (\rm otherwise) ,
  \end{array}\right.  \\
Q &\!\!\!\!=& \!\!\!\! q^{\frac{1}{4}( \#\{ i'|j_{i'}=n\}
    +  \#\{ i'|j_{i'}=n+1\} -1 )   }  .
\label{eq:Q}
\end{eqnarray}
We notice that the matrix
 $\widetilde F^{(n)}$ 
 (resp. $\widetilde E^{(n)}$)
   sends a ket (resp. bra)  vector in the sector
    $m=(m_1,\dots,m_{N+1})$
   to the sector $m^{(n)}
   =(m_1,\dots,m_{n-1},m_n-1,m_{n+1}+1,
   m_{n+2},\dots,m_{N+1})$, i.e. 
$\widetilde F^{(n)} V_m   \subseteq   V_{m^{(n)}}$,
$V^\ast_m \widetilde  E^{(n)} 
 \subseteq   V^\ast_{m^{(n)}} $.
In other words, these matrices change one particle $n$ to $n+1$. 
From the commutation relations 
(\ref{eq:[FM]=[EM]=[KM]=0}), we have 
\begin{equation}
\label{eq:first-conj}
 \widetilde F^{(n)}M_m=M_{m^{(n)}}\widetilde F^{(n)},
 \quad 
 M_m\widetilde E^{(n)}=\widetilde E^{(n)}M_{m^{(n)}},
\end{equation}
which implies 
$\Spec(M_m)\subseteq\Spec(M_{m^{(n)}})$ or 
$\Spec(M_m)\supseteq\Spec(M_{m^{(n)}})$
 according to
$\dim V_m\le\dim V_{m^{(n)}}$
or $\dim V_m\ge\dim V_{m^{(n)}}$, respectively.\footnote{
For the periodic-boundary case,
this spectral inclusion
is not generally satisfied \cite{AKSS}.  }
Here Spec($M_m$) is the multiset of all
eigenvalues of sector $m$,
where the multiplicity of each element
corresponds to the degree
of degeneracy.
%This spectral relation still expected to be satisfied
%in the TASEP case $q=0$. 

For a sector $m=(m_1,\dots,m_{N+1})$,
we define 
\begin{equation}
\label{eq:m^n^mu=}
m^{(n)^\mu}=
(m_1,\dots,m_{n-1},m_n-\mu,m_{n+1}+\mu,
   m_{n+2},\dots,m_{N+1}) ,
\end{equation}
 i.e.  the sector $m^{(n)^\mu}$ is obtained by
changing $\mu $ particles of $n$th class
to  $(n+1)$st-class particles.
The matrix  $(F^{(n)})^\mu$ sends a vector
in $V_m$ to $V_{m^{(n)^\mu}}$,
and one can show by induction
that each element of 
 $(\widetilde F^{(n)})^\mu$ is calculated as
\begin{eqnarray}
 \label{eq:F^mu-element}
   \langle j_1\cdots j_L |
    (\widetilde F^{(n)})^\mu
  | k_1 \cdots k_L \rangle  
= [\mu]!\,  Q^\mu\, 
  q^{\sum_{i\in I} \#  \{ i'<i | j_{i'}=n  \} }, \\
 {\rm if}\ \exists I\subset \{ 1,\dots,L \}
     \ (\#I=\mu)  {\rm \ such \  that \ }  
     \  j_\iota-1=k_\iota=n\ (\iota\in I)
 \  {\rm   and} \ 
 j_\iota = k_\iota (\iota\notin I) ,
\label{eq:condition}
\end{eqnarray}
or 0 otherwise.
Here $[\mu]!$ is the $q$-factorial
  $\prod_{1\le \mu'\le \mu} [\mu']$
with the $q$-integer
 $[\mu']=1+q+\cdots +q^{\mu'-1}$,
 and $Q$ is defined by (\ref{eq:Q}).
Similarly, 
 $(\widetilde E^{(n)})^\mu$ 
 sends a vector in $V^*_m$ to
  $V^*_{m^{(n)^\mu} }$, 
and each element is calculated as
 \begin{eqnarray}
 \label{eq:E^mu-element}
   \langle k_1\cdots k_L |
    (\widetilde E^{(n)})^\mu
  | j_1 \cdots j_L \rangle  
= [\mu]!\,  Q^\mu\, 
 q^{\sum_{i\in I} \#\{i'>i|k_{i'}=n+1\} }
\end{eqnarray} 
if the condition (\ref{eq:condition})
is satisfied, or  0 otherwise.
From equation (\ref{eq:first-conj})
we have 
\begin{equation}
\label{eq:FM=MF,ME=EM}
 (\widetilde F^{(n)})^\mu M_m=M_{m^{(n)^\mu}} (\widetilde F^{(n)})^\mu ,
 \quad 
 M_m (\widetilde E^{(n)})^\mu = (\widetilde E^{(n)})^\mu M_{m^{(n)^\mu}}.
\end{equation}
We call this type of relations
{\it conjugation relation}
and the matrix that satisfies it
{\it conjugation matrix} \cite{AAMP1}.

{\it Comment:}
Since $H^{(n)}$ (\ref{eq:H}) is a symmetric matrix, 
we have $M^{\rm T}=S^2MS^{-2}$.
This means our process satisfies 
the detailed-balance condition,
and the stationary-state probability is expressed as
$P_0(k_1\cdots k_L)=\langle k_1\cdots k_L| S^{-2}
|k_1\cdots k_L\rangle = \frac{1}{Z}
 q^{-\frac{1}{2}\sum_{1\le i<j\le L} {\rm sign} (k_i-k_j)}$. 
This stationary state can be rewritten in terms of 
a matrix product form,
which we will achieve in another way, i.e. 
by using the generalized matrix ansatz,
in section \ref{sec:stationary-state}.

\section{Matrix product interpretation}\label{sec:matrix-product}

In this section, we write the elements
of $(\widetilde F^{(n)})^\mu$ and 
 $(\widetilde E^{(n)})^\mu$
  in a matrix product form.
We first define matrix-valued matrices
 $b^{(n)}$ and  $\bar b^{(n)}$ 
 of size $(N+1)\times(N+1)$ as 
\begin{eqnarray}
\label{eq:b^n}
  b^{(n)} &\!\!\!=& \!\!\! 
  \sum_{1\le x\le N+1 \atop x\neq n} \one|x\rangle\langle x|
 + D|n\rangle\langle n| + A|n+1\rangle\langle n| , \\
 \bar b^{(n)}
  &\!\!\!=& \!\!\! 
  \sum_{1\le x\le N+1 \atop x\neq n+1} \one|x\rangle\langle x|
 + D|n\rangle\langle n+1| + A|n+1\rangle\langle n+1|,
 \end{eqnarray}
where
\begin{eqnarray}
 D=
\left(\begin{array}{cccc}
0  & 1    &           &   \\
   & 0    & 1        &   \\
   &      &  0    &  {\scriptsize\ddots }  \\
   &      &      &    {\scriptsize\ddots }
\end{array}\right),\quad 
A =\left(\begin{array}{cccc}
1  &  &      &   \\
&  q & &       \\
  &  &   q^2  &   \\
  &      &      &   {\scriptsize\ddots }
\end{array}\right),
\label{matrices}
\end{eqnarray}
 satisfying the relations $DA=qAD$.
We also use vectors
 $\brabra w|$ and $|v \ketket$
  defined as
\begin{equation}
\label{eq:w=,v=}
 \brabra w| = (1\ 0\ 0 \ \cdots ),\quad 
|v\ketket = 
\left(\begin{array}{c}
  1  \\ 1 \\  {\scriptsize\vdots }
\end{array}\right),
\end{equation}
which satisfy
$\brabra w| A = \brabra w|$ and 
$D | v \ketket = | v \ketket $.

The elements of
 $(\widetilde F^{(n)})^\mu$ (\ref{eq:F^mu-element}) and 
 $(\widetilde E^{(n)})^\mu$ (\ref{eq:E^mu-element})
can be interpreted as follows:
for configurations
 $ j_1\cdots j_L$ and $k_1 \cdots k_L$
in the sectors $m$ and $m^{(n)^\mu}$
 (\ref{eq:m^n^mu=}),
respectively,  we have
\begin{eqnarray}
\label{eq:F^mu=bbb}
     \langle j_1\cdots j_L | (F^{(n)})^\mu 
  | k_1 \cdots k_L \rangle
  &\!\!\!=& \!\!\!
   [\mu]!\,  Q^\mu\,  \brabra w| b^{(n)}_{j_1k_1}
   \cdots b^{(n)}_{j_Lk_L}| v\ketket,  \\
\label{eq:E^mu=bbb}
     \langle k_1\cdots k_L | (E^{(n)})^\mu 
  | j_1 \cdots j_L \rangle
  &\!\!\!=& \!\!\! [\mu]!\,  Q^\mu\, 
   \brabra w| \bar b^{(n)}_{k_1j_1}
   \cdots  \bar b^{(n)}_{k_Lj_L}| v\ketket,
\end{eqnarray}
where we write
 $b^{(n)}_{jk} =\langle j|b^{(n)} |k\rangle$ 
and  $\bar b^{(n)}_{kj} =\langle k|\bar b^{(n)} |j\rangle$.
Equation  (\ref{eq:F^mu=bbb})
 is  understood as follows.
Let $j_1\cdots j_L$ and $k_1\cdots k_L$
be configurations such that 
\begin{equation}
\label{eq:j=k=n-or}
  j_i = k_i   \quad {\rm or}\quad j_i-1=k_i=n,
\end{equation}
and set 
\begin{equation}
 k_{i_1}=\cdots=k_{i_{\mu+\nu}}=n,\  
 \mu=  \#  I = \#\{  i|  j_i-1=k_i=n  \} ,\ 
  \nu=  \#\{  i|  j_i=k_i=n  \}  .
\end{equation}
Since $b_{j_ik_i}=\mbox{\rm \one} $  for $ k_i\neq n$, we have
$ \brabra w| b^{(n)}_{j_1k_1}\cdots b^{(n)}_{j_Lk_L} |v \ketket
 = \brabra w| b^{(n)}_{j_{i_1}n}\cdots b^{(n)}_{j_{i_{\mu+\nu}}n}
  |v \ketket
$ where $j_{i_\ell}=n,n+1$ and $b^{(n)}_{j_{i_\ell}n} =A,D$.
Noting $\sum_{i\in I}\#\{ i'<i| j_{i'}=n\}$ corresponds 
to how many times
we need to exchange $DA\to AD$ for reordering
 the matrix product
$  b^{(n)}_{j_{i_1}n}\cdots b^{(n)}_{j_{i_{\mu+\nu}}n}  $
to $ \underbrace{A\ \cdots\  A}_{\mu }
\underbrace{D\ \cdots\  D}_{\nu  } $,
we find
\begin{eqnarray}
 \brabra w| b^{(n)}_{j_{i_1}n}\cdots
  b^{(n)}_{j_{i_{\mu+\nu}}n} |v\ketket
 =   q^{ \sum_{i\in I}   \#\{ i'<i | j_{i'}=n  \}  }
\brabra w|   \underbrace{A\ \cdots\  A}_{  \mu }
\underbrace{D\ \cdots\  D}_{  \nu  } |v\ketket 
=  q^{ \sum_{i\in I}   \#\{ i'<i | j_{i'}=n  \}  }. 
\label{eq:MP-element}
\end{eqnarray}
If the configurations do not satisfy
 the condition (\ref{eq:j=k=n-or}), we have 
$\brabra w| b^{(n)}_{j_1k_1}\cdots b^{(n)}_{j_Lk_L}|v\ketket=0$.
Noting this and 
 equations (\ref{eq:F^mu-element})
 and (\ref{eq:MP-element})
 we find the matrix product interpretation (\ref{eq:F^mu=bbb}).
 One can show equation (\ref{eq:E^mu=bbb}) 
in the same way.
Defining the matrices
\begin{eqnarray}\label{eq:psi}
  \langle j_1\cdots j_L |\psi_{m^{(n)^\mu},m}
  | k_1\cdots k_L \rangle 
  &\!\!\!=& \!\!\!  \brabra w| b^{(n)}_{j_1k_1}\cdots b^{(n)}_{j_Lk_L}     |v \ketket ,\\
  \langle k_1\cdots k_L |\varphi_{m,m^{(n)^\mu}}
  | j_1\cdots j_L \rangle 
  &\!\!\!=& \!\!\! \brabra w| \bar b^{(n)}_{k_1j_1}\cdots\bar b^{(n)}_{k_Lj_L}|v \ketket ,
\label{eq:phi}
\end{eqnarray}
we rewrite the conjugation relation 
(\ref{eq:FM=MF,ME=EM}) as
\begin{equation}
\label{eq:psiM=Mpsi,Mphi=phiM}
\psi_{m^{(n)^\mu},m} M_m
=M_{m^{(n)^\mu}}\psi_{m^{(n)^\mu},m},
 \quad 
 M_m\varphi_{m,m^{(n)^\mu}}
  = \varphi_{m,m^{(n)^\mu}} M_{m^{(n)^\mu}}.
\end{equation}
In what follows, we show these
relations in another way starting 
from local relations different from (\ref{eq:[D,H]=0s}).

Using the relation $DA=qAD$,
one can check that the tensor products
 $ b^{(n)}\ot b^{(n)}$ and 
 $\bar b^{(n)}\ot\bar b^{(n)}$
commute  with the local Markov matrix:
\begin{eqnarray}\label{eq:[bb,M]=0}
 \left[ b^{(n)}\ot b^{(n)} , M^{(N)}_{\rm Loc} \right] 
=\left[ \bar b^{(n)}\ot \bar b^{(n)}
 , M^{(N)}_{\rm Loc} \right]  = 0.
\end{eqnarray}
This leads to a global commutation relation
\begin{eqnarray}
  \left[ \left(b^{(n)}\right)^{\ot L}, M^{(N)}\right] 
  =\left[ \left(\bar b^{(n)}\right)^{\ot L},
   M^{(N)}\right] 
  = 0.
\end{eqnarray}
Bookending each element
  between  vectors $\brabra w|$ and $|v \ketket$
  (\ref{eq:w=,v=}),
  we obtain  (scalar-valued) matrices 
  of size $(N+1)^L\times(N+1)^L$ 
\begin{equation}
\Psi^{(n)} =\brabra w|\left(b^{(n)}\right)^{\ot L} |v\ketket, 
\quad
 \Phi^{(n)} =\brabra w|
 \left(\bar b^{(n)}\right)^{\ot L} |v  \ketket
\end{equation}
which satisfy
\begin{eqnarray}
\label{eq:[Psi,M]=0}
  \left[\Psi^{(n)}, M^{(N)}  \right] =
  \left[\Phi^{(n)}, M^{(N)}  \right] = 0.
\end{eqnarray}
Since  $\psi_{m,m^{(n)^\mu}}$ (\ref{eq:psi})
 and $\varphi_{m^{(n)^\mu},m}$ (\ref{eq:phi})
are submatrices of $\Psi^{(n)}$ and $\Phi^{(n)}$,
respectively, the commutation relations
 (\ref{eq:[Psi,M]=0}) lead to
  the conjugation relations
   (\ref{eq:psiM=Mpsi,Mphi=phiM}).

\section{Relation between dynamics of 
different values of $N$}\label{sec:diffN}

In this section, we construct a matrix 
which intertwines dynamics of 
an $N$-species sector $m=(m_1,\cdots,m_{N+1})$ 
and an $(N- 1)$-species 
 sector $m'=(m_1,\dots,m_{n-1},m_n+m_{n+1},
 m_{n+2},\cdots,m_{N+1})$.
The dynamics of the sector $m'$ 
is essentially same as that of the sector 
$ \overline{m} =(m_1,\dots,m_{n-1},
m_n+m_{n+1}, 0,m_{n+2},\cdots, m_{N+1})$
by regarding particles of $x$th class  ($x\ge n+2$)
as that of $(x\!  -\! 1)$st class.
(Note that $ \overline{m}^{(n)^{m_{n+1}} } = m$.) 
Thus we have already known that
there exit  matrices
 $ \psi_{mm'}$ and $ \varphi_{m'm}$ that satisfy
 conjugation relations 
$ M_m \psi_{mm'} =  \psi_{mm'} M_{m'} $ and 
 $\varphi_{m'm}M_m = M_{m'}\varphi_{m'm}$.
 
Now we restrict our consideration to
 sectors $(m_1,\dots,m_{N+1} )$ such that 
  $m_i>0$ for all $i$ ({\it basic sector}), and
 introduce alternative labeling for basic sectors \cite{AKSS}:
\begin{equation}\label{eq:m<->s}
 m= (m_1,\dots,m_{N+1})\  \leftrightarrow\ 
  \frs = \{ s_1 , \dots  , s_N \}\quad
  (1\le s_1<\cdots < s_N \le L-1 )
\end{equation}
with the correspondence $s_i=m_1+\cdots + m_i $,
or equivalently $m_i=s_i-s_{i-1}$ ($s_0=0,s_{N+1}=L$).
In particular, for the zero-species sector,
we use the labeling 
\begin{equation}
  (L) \  \leftrightarrow\   \emptyset .
\end{equation}
According to this correspondence,
the sector $m' $ is labeled by $\frs\setminus \{s_n\}$.
We write $V_\frs=V_m$ and $M_\frs=M_m$
for $\frs\leftrightarrow m$.

First we  construct
 $\psi_{mm'} = \psi_{\frs ,\frs\setminus \{s_n\}}$
 that satisfies the conjugation relation
 by starting with the following
matrix-valued matrix $a^{(N,n)}$
of size $(N+1)\times N$,
  a ``degenerated version'' of the matrix
  $b^{(n)}$ (\ref{eq:b^n}):
\begin{eqnarray}
\label{eq:a=}
   a^{(N,n)}  
= \sum_{1\le x\le N+1 \atop x\neq n,n+1} \one 
| x \rangle\langle \chi_n(x) |
  + D |n\rangle\langle n|  + A |n+1\rangle\langle n| ,
\end{eqnarray}
 where 
\begin{equation}
  \chi_n(x) 
  = \left\{\begin{array}{ll}
    x &   (x\le n),   \\   x-1 &  (x>n) .
  \end{array}\right.
\end{equation}

By using the relation $DA=qAD$,
one can check that 
this matrix satisfies the following  relation,
 a ``degenerated version'' of the commutation relation
 (\ref{eq:[bb,M]=0}):
\begin{eqnarray}\label{eq:hat-relation}
M^{(N)}_{\rm Loc}(a^{(N,n)}\otimes a^{(N,n)})-
(a^{(N,n)}\otimes a^{(N,n)})M^{(N-1)}_{\rm Loc}= 0.
\end{eqnarray}
From this relation,
 the $L$-fold tensor product of $a^{(N,n)}$ satisfies
\begin{eqnarray}
\sum_{1\le i\le L-1}\left(M^{(N)}_{\rm Loc}\right)_{i,i+1}
\left(a^{(N,n)}\right)^{\ot L}-\left(a^{(N,n)}\right)^{\ot L}
\sum_{1\le i\le L-1}\left(M^{(N-1)}_{\rm Loc}\right)_{i,i+1} =0.
\end{eqnarray}
Noting that the summations of local Markov matrices
   are total Markov matrices, we get
\begin{eqnarray}\label{MTr=TrM}
  M^{(N)} \Psi^{(N,n)}  =  \Psi^{(N,n)} M^{(N-1)},
\end{eqnarray}
where $\Psi^{(N,n)}=\brabra w | \left(a^{(N,n)}\right)^{\ot L}  |v\ketket$.
Thus we find the  submatrix 
$ \psi_{\frs ,\frs\setminus \{s_n\}} $ of $ \Psi^{(N,n)} $,
i.e.
\begin{equation}
 \psi_{\frs ,\frs\setminus \{s_n\}}:
V_{\frs\setminus \{s_n\}} \to V_\frs,
\quad
\label{eq:phi-element}
\langle j_1\cdots j_L |  \psi_{\frs ,\frs\setminus \{s_n\}}   
| k_1\cdots k_L \rangle
= \brabra w| a^{(N,n)}_{j_1k_1}\cdots
   a^{(N,n)}_{j_Lk_L}|v\ketket,
\end{equation}
with $a^{(N,n)}_{jk}=\langle j| a^{(N,n)} |k\rangle $,
 satisfies the conjugation relation
\begin{equation}
\label{eq:a-goal}
  M_\frs \psi_{\frs,\frs\setminus\{s_n\}}
  = \psi_{\frs,\frs\setminus\{s_n\}} M_{\frs\setminus\{s_n\}}  .
\end{equation}  
Each element (\ref{eq:phi-element}) of the conjugation matrix
$\psi_{\frs ,\frs\setminus \{s_n\}}$
 becomes  0 or a power of $q$ 
with its exponent corresponding to how many times
we need to exchange $DA\to AD$
for reordering the matrix product 
to $ \underbrace{A\ \cdots\  A}_{s_{n+1}-s_n}
\underbrace{D\ \cdots\  D}_{s_n-s_{n-1} } \,  $:
\begin{eqnarray}
   \langle j_1\cdots j_L | 
  \psi_{\frs ,\frs\setminus \{s_n\}}
   |k_1\cdots k_L \rangle =
\left\{\begin{array}{ll}
  \exp \Big(\ln q \displaystyle\sum_{i:j_i=n+1}
   \#\{ i'<i | j_{i'}=n  \}   \Big) 
   &  (\chi_n(j_i)=k_i\ (\forall i)  ), \\
   0  & ({\rm otherwise})  .
\end{array}\right.
\label{eq:element-explicit}
\end{eqnarray}
For example, for  the sectors
 $\frs=\{ 1,3,5 \} \leftrightarrow (1,2,2,1)$
  and $\frs\setminus\{s_2\} = \{1,5\} \leftrightarrow (1,4,1)$
   with $L=6$,
\begin{eqnarray}
\label{eq:DA->AD-example}
& & \nonumber
\langle 123423 | \psi_{\frs,\frs\setminus\{s_2\} } | 122322 \rangle
=\brabra w| \one DA \one  DA |v\ketket
=\brabra w| DADA |v\ketket
\\
& & 
=q\brabra w| DAAD |v\ketket
=q^2\brabra w| ADAD |v\ketket
=q^3\brabra w| AADD |v\ketket=q^3.
\end{eqnarray}

The generalized matrix (product) ansatz, 
i.e. the procedure 
(\ref{eq:hat-relation})-(\ref{eq:a-goal}), 
 was introduced in \cite{AAMP1}
 for the periodic boundary condition.
There the right-hand side of (\ref{eq:hat-relation})
is replaced as
\begin{eqnarray}
 M^{(N)}_{\rm Loc}(a^{(N,n)}\otimes a^{(N,n)})-
(a^{(N,n)}\otimes a^{(N,n)})M^{(N-1)}_{\rm Loc}  
= a^{(N,n)} \otimes\widehat{a}^{(N,n)}-\widehat{a}^{(N,n)}\otimes a^{(N,n)}
\label{eq:hat-origi}
\end{eqnarray}
 with an auxiliary matrix $\widehat{a}^{(N,n)}$,
 and the conjugation matrix is constructed by
 taking the trace
\begin{eqnarray}
\label{eq:psi-period}
\langle j_1\cdots j_L| \psi_{\frs ,\frs\setminus \{s_n\} }
   |k_1\cdots k_L\rangle
= \Tr \left[a^{(N,n)}_{j_1k_1}
\cdots a^{(N,n)}_{j_Lk_L} \right] .
\end{eqnarray}
(We set $\widehat{a}^{(N,n)}=0$ in our case.)
The families of representations for
 the {\it hat relation} (\ref{eq:hat-origi})
 found in \cite{AAMP1,AAMP2}
are more complicated than our case.
For example, for $(N,n)=(3,1)$,
our representation (\ref{eq:a=}) is
\begin{equation}
\label{eq:31our-solution}
a^{(3,1)}=
\bordermatrix{
                 &{}_{1}  & {}_{2}     & {}_{3}    \cr
 \scriptstyle{1} & D & 0 & 0 \cr
 \scriptstyle{2} & A  & 0  &0 \cr
 \scriptstyle{3} & 0  & \one  & 0   \cr
 \scriptstyle{4} & 0  & 0  & \one   } 
 \quad ({\rm reflective}).
\end{equation}
On the other hand, the representation found in
 \cite{AAMP1,AAMP2} is 
\begin{eqnarray}
\label{eq:31periodic-solution}
a^{(3,1)}=
\bordermatrix{
                 &{}_{1}  & {}_{2}     & {}_{3}    \cr
 \scriptstyle{1}
  & \one\ot\one & \del\ot\one  & \one\ot\del \cr
 \scriptstyle{2} &
  A\ot A & 0 & 0  \cr
 \scriptstyle{3} & 
 \eps\ot A & \one\ot A & 0   \cr
 \scriptstyle{4} &
 \one\ot\eps & \del\ot \eps & \one\ot\one  }
 \quad ({\rm periodic}),
\end{eqnarray}  
where
\begin{equation}
 \delta =
\left(\begin{array}{cccc}
  0   &  c_1   &      &   \\
       &   0  &   \!\! c_2 &    \\[-2mm]
       &      &   \!\! 0  & \!\! \ddots  \\[-2mm]
       &      &      &  \!\! \ddots
\end{array}\!\!\!\right),\ 
 \epsilon=
\left(\begin{array}{cccc}
   0   & &      &   \\
   c_1 &   0  &  &   \\
       & c_2  & \!\! 0  & \\[-2mm]
      &      & \!\!\ddots &   \!\!\ddots
\end{array}\!\!\!\right),\ 
c_\nu=\sqrt{1-q^\nu},
\end{equation}  
and $\widehat{a}^{(N,n)}$ needs to be chosen as 
${\rm diag}(1,q,q,q)a^{(3,1)}$.
Note that the number of tensor products
of each element in the solution for the periodic case
increases as $N$ increases.
On the other hand, the solution (\ref{eq:a=})   
 does not contain a tensor product.
 
The hat relation (\ref{eq:hat-origi}) is independent
  from boundary conditions.
However, whether a representation
 for the algebra defined by the hat relation 
 is practical
  (i.e. whether a representation
  allows us to construct a nontrivial conjugation matrix)
  depends on boundary conditions.
For example, the matrix $\psi_{\frs ,\frs\setminus \{s_n\}}$
  defined by (\ref{eq:psi-period}) 
 with the representation (\ref{eq:31our-solution}) is 0.

In the same way, we can construct 
the restricted version  for $\varphi$ (\ref{eq:phi}),
starting with the  matrix 
$\bar a^{(N,n)}  =
 \sum_{1\le x\le N+1 \atop x\neq n,n+1} \one 
| \chi_n(x) \rangle\langle x |
  + D |n\rangle\langle n+1| $.
The matrix $\varphi_{\frs\setminus\{s_n\}\frs}$
defined by
$\langle k_1\cdots k_L|\varphi_{\frs\setminus\{s_n\}\frs} |j_1\cdots j_L\rangle
=\brabra w|\bar a_{k_1j_1}^{(N,n)}\cdots
\bar a_{k_Lj_L}^{(N,n)}|v \ketket $
with $\bar a_{kj}^{(N,n)}
= \langle k|\bar a ^{(N,n)}|j\rangle$ 
satisfies the conjugation relation 
$\varphi_{\frs\setminus\{s_n\}\frs}M_\frs=
M_{\frs\setminus\{s_n\}} 
   \varphi_{\frs\setminus\{s_n\}\frs}$.
This matrix has indeed trivial elements
\begin{equation}
     \varphi_{\frs\setminus \{s_n\},\frs} =
     \sum    |\chi_n( j_1) \cdots \chi_n( j_L)   \rangle
      \langle   j_1\cdots j_L |  ,
\end{equation}
where the summation runs over all the configuration
in the sector $\frs$.
For ket vectors, this matrix ``identifies''
  $n$th and $(n+1)$st class particles 
  as a same class \cite{AKSS}.
The matrix $\psi_{\frs ,\frs\setminus \{s_n\}}$
(\ref{eq:phi-element}) 
 sends any basis vector $|k_1\cdots k_L\rangle \in V_{\frs\setminus \{s_n\}}$
  to vectors in $V_{\frs}$
  such that each particle $k_i$ keeps its position $i$,
   $(s_{n+1}-s_n)$ particles of $n$th class
  are changed to $(n+1)$st-class particles
  (the rest of $(s_n-s_{n-1})$ particles of 
  $n$th-class  are unchanged),
   and $\nu$th-class particles ($\nu>n$)
  are changed to $(\nu+1)$st-class particles.
Such vectors are restored to $|k_1\cdots k_L\rangle$ by
the identification matrix
  $\varphi_{\frs\setminus\{s_n\} ,\frs}$:
\begin{eqnarray}
\nonumber
  \varphi_{ \frs\setminus \{s_n\},\frs}
   \psi_{\frs ,\frs\setminus \{s_n\} }|k_1\cdots k_L\rangle
  =   \varphi_{\frs\setminus \{s_n\},\frs}
   \sum_{j_1\cdots j_L: \atop \chi_n( j_i)= k_i\, (\forall i)}
  \brabra w| a_{j_1k_1} \cdots a_{j_Lk_L}
    |v\ketket 
  |j_1\cdots j_L\rangle  \\
  =\sum_{j_1\cdots j_L: \atop \chi_n(j_i)= k_i\, (\forall i)}
  \brabra w| a_{j_1k_1} \cdots a_{j_Lk_L}    |v\ketket
  |k_1\cdots k_L\rangle
  =:  C  |k_1\cdots k_L\rangle.
\label{eq:=|>sum=C|>}
\end{eqnarray}
Actually the constant $C$ is independent of the configuration  $k_1\cdots k_L$
and one can show
\begin{equation}
\label{eq:C=}
  C= \sum_{U_i=D,A \atop 
  \#\{i| U_i=D  \}  = s_n-s_{n-1}  }
  \brabra w|  U_1 \cdots
   U_{s_{n+1}-s_{n-1} }
   |v\ketket
  =\frac{ [ s_{n+1} -s_{n-1} ]! }
  { [s_n-s_{n-1} ]! [s_{n+1} -s_{n} ]! }   .
\end{equation}
Thus we have
\begin{eqnarray}
\label{eq:phipsi=}
  \varphi_{ \frs\setminus \{s_n\},\frs} \psi_{\frs ,\frs\setminus \{s_n\} }
 =\frac{ [ s_{n+1} -s_{n-1} ]! }
  { [s_n-s_{n-1} ]! [s_{n+1} -s_{n} ]! }
     {\rm Id}_{\frs\setminus\{s_n\}} ,
\end{eqnarray}
where Id is the identity matrix.
The injectivity of $\psi_{\frs ,\frs\setminus \{s_n\}}$
follows from this relation, and we have
the inclusion relation
\begin{equation}
  \label{eq:inc2}
  \Spec{M_\frs}
  \supset
  \Spec{M_{ \frs\setminus \{s_n\}}     }.
\end{equation}

Now we turn to the construction
of the conjugation matrix
  between Markov matrices
  of $N$- and $N'$-species sectors
  ($N-N' = u >0 $).
Let us consider the local relation,
which we also call hat relation,
\begin{eqnarray}\label{eq:big-hat-relation}
M^{(N)}_{\rm Loc}(\mathcal X\otimes\mathcal X)-
(\mathcal X\otimes \mathcal X)
M^{(N')}_{\rm Loc}= 0.
\end{eqnarray}
We have already known a family of solutions to this relation:
\begin{eqnarray}
  \mathcal X= a^{(N,n_1) } \star\cdots\star  a^{(N'+1,n_u )} \quad
  (1\le n_\ell \le N-\ell+1  ),
\end{eqnarray}
where the symbol $\star$ denotes 
 the product
 $Q\star R = \left\{ \sum_jQ_{ij}\ot R_{jk} \right\}_{ik}$
 for matrix-valued matrices
  $Q=\{Q_{ij}\}_{ij}$ and $R=\{R_{ij}\}_{ij}$.
For example,
\begin{eqnarray}
\mathcal X=  a^{(3,2)}\star a^{(2,1)}
=\left(\begin{array}{ccc}
   \one & 0 & 0 \\
     0  & D & 0 \\
     0  & A & 0 \\
     0  & 0 & \one 
 \end{array}\right)
\star \left(\begin{array}{cc}
      D & 0 \\
      A & 0 \\
      0 & \one 
 \end{array}\right)
=\left(\begin{array}{ccc}
   \one\ot D & 0 \\
     D \ot A & 0 \\
     A \ot A & 0 \\
     0  &  \one \ot \one
 \end{array}\right) 
 \end{eqnarray}
 is a solution  to 
 $ M^{(3)}_{\rm Loc} ( \mathcal X \ot \mathcal X )
 = (\mathcal X\ot\mathcal X) M^{(1)}_{\rm Loc} $.

Bookending each element of
 $\mathcal X^{\ot L}$
  between $\brabra w|^{\ot u}$
  and $|v\ketket^{\ot u }$, we obtain
\begin{eqnarray}
\label{eq:big-Psi}
\Psi
=\brabra w|^{\ot u}\mathcal X^{\ot L}
  |v\ketket^{\ot u}
= \Psi^{(N,n_1 ) }\cdots \Psi^{(N'+1,n_u )} ,
\end{eqnarray}
which satisfies
\begin{eqnarray}
   M^{(N)}\Psi = \Psi M^{(N')}.
\end{eqnarray}
The matrix $\Psi$ sends a vector of an $N'$-species sector
to an $N$-species sector,
via $(N'+1)$-species $\to$
 $(N'+2)$-species 
$\to \ \cdots\  \to $  $(N-1)$-species.
Each index $n_\ell $ specifies
  which class of particles splits 
  in sending an $(N-\ell)$-species vector
  to an $(N-\ell+1)$-species sector.

For the $N$-species sector $\frs=\{s_1<\cdots <s_N\}$
and the $N'$-species sector 
 $\frt=\frs\setminus  \{ s_{\nu_1} ,\dots , s_{\nu_u}   \}$
($u=N-N'$),
we have the conjugation matrix 
\begin{equation}
  \psi_{\frs\frt}  =
  \psi_{\frs,\frs\setminus\{s_{\nu_1}  \} }
  \psi_{\frs\setminus\{s_{\nu_1}  \},\frs\setminus\{s_{\nu_1},s_{\nu_2}  \} }
  \cdots   \psi_{ \frt\, \cup\{s_{\nu_u}\} ,\frt }
\end{equation}
satisfying 
\begin{equation}
 M_{\frs} \psi_{\frs\frt}  
 = \psi_{\frs\frt}  M_{\frt}\,  .
\end{equation}
This is a submatrix of $\Psi$ (\ref{eq:big-Psi})
with the choice
 $n_\ell = \nu_\ell - \# \{ z | z<\ell , \nu_z<\nu_\ell \} $,
 but indeed independent of the choice.
It is enough to show the simplest case $u=2$:
\begin{equation}
\label{eq:psipsi=psipsi}
\mbox{ ``commutativity'':}\quad
   \psi_{\frs,\frs\setminus\{ s_\mu \} }
   \psi_{\frs\setminus\{ s_\mu \},\frs\setminus\{ s_\mu,s_\nu \}}
 = \psi_{\frs,\frs\setminus\{ s_{\nu} \} }
   \psi_{\frs\setminus\{ s_\nu \},\frs\setminus\{ s_\mu,s_\nu \}} .
\end{equation}
We suppose $\mu<\nu$,
and use the explicit expression (\ref{eq:element-explicit}).
The choices of $n_\ell\, $s
for the left- and right-hand sides of (\ref{eq:psipsi=psipsi}) 
are  $(n_1,n_2) = (\mu ,\nu-1)$ and $(\nu,\mu)$, respectively.
By the expression (\ref{eq:element-explicit}),
both sides are calculated as 
\begin{eqnarray}
& & \langle j_1\cdots j_L | \psi_{\frs,\frs\setminus\{ s_\mu \} }
   \psi_{\frs\setminus\{ s_\mu \},\frs\setminus\{ s_\mu,s_\nu \}}
  |k_1\cdots k_L \rangle =  \\
\nonumber
& & \left\{\begin{array}{ll}
\exp \Big(\ln q \Big(
\displaystyle\sum_{i:j_i=\mu+1}\#\{ i'<i | j_{i'}=\mu\}
+\displaystyle\sum_{i:\chi_\mu(j_i)=\nu}\#\{ i'<i | \chi_\mu(j_{i'})=\nu-1\}
     \Big)\Big)
  &  \!\!\!\!\!\!\left(\begin{array}{c}
  \chi_{\nu-1}(\chi_\mu(j_i)) \\
   =k_i \ (\forall i) \end{array}\right), \\
   0 &  ({\rm otherwise}),
\end{array}\right. \\
& &  \langle j_1\cdots j_L | \psi_{\frs,\frs\setminus\{ s_\nu \} }
   \psi_{\frs\setminus\{ s_\nu \},\frs\setminus\{ s_\mu,s_\nu \}}
  |k_1\cdots k_L \rangle = \\
& &  \left\{\begin{array}{ll}
\exp \Big(\ln q \Big(
\displaystyle\sum_{i:j_i=\nu+1}\#\{ i'<i | j_{i'}=\nu\}
+\displaystyle\sum_{i:\chi_\nu(j_i)=\mu+1}\#\{ i'<i | \chi_\nu(j_{i'})=\mu\}
     \Big)\Big)
  &  \!\!\!\!\!\!\left(\begin{array}{c}
   \chi_{\mu}(\chi_\nu(j_i)) \\
  =k_i \ (\forall i) 
  \end{array}\right),  \\
   0 &  ({\rm otherwise}),
\end{array}\right.
\nonumber
\end{eqnarray}
which are equal.

\section{Stationary state}\label{sec:stationary-state}

The zero-species sector
  $\emptyset$ 
consists only of the configuration $1\cdots 1$.
Since $M^{(0)}_{\rm Loc}=0$,
the hat relation (\ref{eq:big-hat-relation})
with $N'=0$ becomes 
$M^{(N)}_{\rm Loc}
(\mathcal X\otimes\mathcal X)=0$.
A solution to this hat relation is given as
 $\mathcal X=a^{(N,n_1)}\star a^{(N-1,n_2 ) }
 \star\cdots\star a^{(1,1)}$,
 and thus we obtain a stationary state
 in the matrix product form \cite{ADR,BE,DEHP}:
 the probability $P(j_1\cdots j_L)$
 of finding a configuration $ j_1\cdots j_L $
 can be expressed as
\begin{eqnarray}\label{mpr}
   P(j_1\cdots j_L) =  \frac{1}{Z}
    \brabra  w|^{\ot N}  X_{j_1}\cdots X_{j_L}  |v\ketket^{\ot N},
\end{eqnarray}
where
\begin{eqnarray}
\label{eq:decomp}
X_i= \langle i |
a^{(N,n_1)}\star a^{(N-1,n_2 ) }\star\cdots\star a^{(1,1)}
|1\rangle   
\end{eqnarray}
satisfying $X_\alpha X_\beta = q X_\beta X_\alpha$
($\alpha < \beta $)
\footnote{
The stationary state for the
multis-pecies ASEP
with the general hopping rule
$\alpha\beta\to\beta\alpha$
(rate $\Gamma_{\alpha\beta}$)
has also the matrix product form 
with the algebra
$ X_\alpha X_\beta
= \frac{\Gamma_{\beta\alpha}}{\Gamma_{\alpha\beta}} 
 X_\beta X_\alpha $,
see \cite{ADR} for $N=3$. 
We need a $\frac{N(N+1)}{2}$-fold tensor product in the representation for this algebra,
and the decomposition structure
 (\ref{eq:decomp}) no longer exists.}.
This implies that the system satisfies 
the detailed-balance condition
as we commented in section \ref{sec:symmetry}.

The stationary states in the periodic boundary condition
can also be written in the matrix product form
   \cite{AAMP1,PEM}.
However the representation 
 for the matrices ($X_i$'s)  in our case
 is much simpler than that of the periodic-boundary case.
For example, for $N=3$
with the choice $n_2=n_3=1$, we have
\begin{eqnarray}
a^{(3,1)}\star a^{(2,1)}\star a^{(1,1)}
&\!\!\!=& \!\!\!
\left(\begin{array}{c}
  X_1 \\  X_2 \\  X_3 \\ X_4
\end{array}\right)
= \left(\begin{array}{c}
  D\ot D\ot D  \\
    A\ot  D\ot D \\
   \one\ot  A\ot D \\ \one\ot\one\ot A
\end{array}\right)  \ ({\rm reflective}) ,  \\
a^{(3,1)}\star a^{(2,1)}\star a^{(1,1)}
&\!\!\!=& \!\!\!
\left(\begin{array}{c}
  \one\ot\one\ot(\one+\del)
 +\del\ot\one\ot A
 +\one\ot\del\ot(\eps+\one)  \\ 
  A\ot  A\ot(\one+\del) \\
   \eps\ot A\ot(\one+\del)
 +\one\ot A\ot A \\
  \one\ot\eps\ot(\one+\del)
 +\del\ot\eps\ot A
 +\one\ot\one\ot(\eps+\one)  \end{array}\right) 
  \ ({\rm periodic}) .
\end{eqnarray}

The stationary state $ |P_0\rangle_\frs $
of each sector $\frs=\{ s_1<\cdots <s_N\}$
 is given by a product of conjugation matrices 
 and $|1\cdots 1\rangle$ as 
\begin{equation}
\label{eq:stationary}
   |P_0\rangle_\frs
   =\frac{1}{Z}
    \psi_{\frs, \frs\setminus\{s_{\nu_1}\}} \cdots
     \psi_{ \{s_{\nu_N}  \},\emptyset }
    |1\cdots 1\rangle 
\end{equation}
with $\{\nu_1,\dots,\nu_N\} = \{1,\dots,N\}$.
In other words,
the stationary state in sector $\frs$
is constructed by 
transferring the zero-species vector 
to $V_\frs$ via 
sectors $\{s_{\nu_N}  \}
\to \{s_{\nu_N} ,s_{\nu_{N-1}} \}
\to \cdots \to \frs\setminus \{s_{\nu_1}\} $.
Note that the stationary state (\ref{eq:stationary}) 
is indeed independent of the choice of
intermediate sectors
 (i.e. independent of the choice of
  $\nu_\ell\, $s.)
Noting equation (\ref{eq:phipsi=}),
we find $Z$ of the  sector
 $\frs=\{s_1<\cdots<s_N \}$
in the general case is given by the $q$-multinomial
\begin{eqnarray}
& & \nonumber
   Z = \sum_{ j_1\cdots j_L\atop {\rm in \  sector }\ \frs } P(j_1\cdots j_L)
   = \sum_{ j_1\cdots j_L\atop {\rm in \  sector }\ \frs }
       \langle j_1\cdots j_L |P_0\rangle_\frs \\
& & \quad
   = \langle 1\cdots 1| 
   \varphi_{ \emptyset \{s_N   \} } \cdots 
   \varphi_{ \frs\setminus\{ s_1 \},\frs} \ 
   \psi_{  \frs ,  \frs\setminus\{ s_1  \} } 
   \cdots \psi_{ \{s_N  \} \emptyset  }
   |1\cdots 1\rangle \\
& & \quad
    =\frac{[L]!}{ [s_1-s_0]! [s_2-s_1]! \cdots  [ s_{N+1} -s_N ]! } 
   \quad ( s_0=0,\, s_{N+1} = L ) ,
   \nonumber
\end{eqnarray}
which is the dimension of the sector $\frs$ 
for $q=1$.

For the TASEP case $q=0$,
 a unique configuration can be realized in the stationary state
 since $X_\alpha X_\beta=0\ (\alpha<\beta)$.
That is, the stationary state in each sector is
   an absorbing state,
 where all the particles stay
 in the descending order
$   j_1 \cdots j_L \ 
(j_i\ge j_{i+1}\ {\rm for}\  \forall i)$.
On the other hand,
for the SSEP case $q=1$,
all the possible configurations 
are realized with an equal probability.

Now we turn to the relaxation
to the stationary state,
where the relaxation time $\tau$
 is characterized by
 the largest non-zero eigenvalues $E$
 as $\tau =-{\rm Re}\, E^{-1}$.
 (The largest eigenvalue is indeed 0,
 which corresponds to the stationary state.)
We first consider the simplest sector 
  $\{L-1\}\leftrightarrow  (L-1,1)$, i.e.
   $L-1$ partiles and 1 vacancy.
The spectrum of $M_{ \{L-1\} }$ is given by 
$     \Spec(    M_{ \{L-1\} }    )  = \{0\} \cup
\{  -(1+q)  + 2\sqrt{q}\cos \frac{k\pi}{L} |
 k=1,\dots,L-1  \}$.\footnote{
This can be easily derived by the Bethe ansatz,
and eigenvectors (except the stationary state)
are given by
$  |P_k\rangle_{\{ L-1 \}} = 
\sum_{1\le i\le L-1}  \left(\lambda^i  + 
\frac{\lambda(1-q\lambda)}{1-\lambda}
  (q\lambda)^{-i}\right)  
   | 1\!\cdots\! 1  \ue{i{\rm th}} {2}
   1\!\cdots\! 1  \rangle $
with $\lambda = \frac{1}{\sqrt{q}}
{\rm e}^{{\rm i} k/L}$
 $(k=1,\dots,L-1)$.    }
In particular the largest non-zero eigenvalue is 
\begin{equation}
E=-(1+q)  + 2\sqrt{q}\cos \frac{\pi}{L}.
\label{eq:E}
\end{equation}
The spectrum of 
 the general one-{\it species} sector 
$\{s_1\}\leftrightarrow (s_1,L-s_1)$ 
($1<s_1  <L$)
contains  that of 
the one-{\it vacancy} sector $\{L-1\}$. 
Furthermore we expect that
 the largest non-zero eigenvalue of the sector  
$\{s_1\}$ is equal to 
that of $\{L-1\}$ (\ref{eq:E}),
which can be checked in small systems.
This observation implies that  the relaxation time 
behaves as \cite{Sa}
\begin{equation}
\label{eq:tau-simeq}
\tau\simeq (1-\sqrt{q})^{-2} \ 
(q<1),\  \frac{L^2}{\pi}\ (q=1),
\end{equation}
as $L\to\infty$.
The ``fist excited state'' $ | P_1 \rangle_{ \{s_N\} }$,
i.e. the eigenvector corresponding
  to the largest non-zero eigenvalue 
  of the sector $   \{s_N\} $ can be written as 
$  | P_1 \rangle_{ \{s_N\} }
  =     \psi_{ \{s_N\}  \{L-1\}   }
     |    P_1 \rangle _{ \{L-1\}   }$
where $ \psi_{ \{s_N\}  \{L-1\}   }$
is constructed as (\ref{eq:psi}).
Recall that the spectrum of the  general 
multi-species sector $\frs=\{s_1<\cdots<s_N\}$
also contains that of the sector 
$\{s_N\}$, see (\ref{eq:inc2}).
Again we expect 
the largest non-zero eigenvalue of the sector $\frs$
is identical to (\ref{eq:E}),
which can be checked 
 for sectors with small dimensions.
This implies the same behavior 
of the relaxation time (\ref{eq:tau-simeq})
as for the one-species case.
The corresponding eigenvector 
of the sector $\frs$ also has the form
$  | P_1 \rangle _\frs  
   = \psi_{\frs, \frs\setminus\{s_{\nu_1}\}} \cdots
     \psi_{ \{s_{N-1},s_N\},\{s_N  \}    }
     |    P_1 \rangle _{ \{ s_N \}  } .$
In contrast to our case,
the relaxation time of the multi-species ASEP
behaves as
\begin{equation}
\label{eq:tau-periodic}
   \tau\sim  L^{\frac{3}{2}} \ 
(q<1),\   L^2   \ (q=1) 
\end{equation}
in the periodic boundary condition \cite{AKSS,K}.
Comparing the behaviors (\ref{eq:tau-simeq})
and (\ref{eq:tau-periodic}),
we notice that the boundary condition
plays an important role.

\section{Summary}\label{sec:summary}

We investigated a multi-species generalization
of the ASEP with reflective boundaries.
We found the symmetry of the Markov matrix
can be interpreted as a matrix product form,
 constructing conjugation matrices 
  (\ref{eq:psi}) and  (\ref{eq:phi})
  which intertwine  Markov matrices
   of different sectors.
We showed that 
the conjugation relations follow  from 
the local commutation relations (\ref{eq:[bb,M]=0}).
We also considered
relations between Markov matrices
 of  different values of $N$.
We constructed a conjugation matrix
connecting dynamics of a simpler system  
and  a more complex system 
by using solutions to
 the hat relation (\ref{eq:hat-origi}).
We saw that the stationary state 
can be written in a product of conjugation matrices,
and the first excited state can be obtained by
multiplying that of one-species sector
by a product of conjugation matrices.
These properties are also true
 in the periodic boundary condition.
(However the behaviors of the relaxation time
to the stationary states are different
in these two boundary conditions
in general.)

It is remarkable that 
there exist several solutions
(representations) $a^{(N,n)}$ to 
the hat relation (\ref{eq:hat-origi})
which are suitable either the reflective 
boundary condition or the periodic one
 \cite{AAMP1,AAMP2},
and the choice of the hat matrix
 $\widehat a^{(N,n)}$
have to be changed according to the boundary conditions.
At present we do not have a systematic way
 to find  appropriate representations
 as well as appropriate hat matrices.
Another interesting study will be 
applying (or generalizing) our method
to the system with injection and extraction of particles
\cite{BE,DEHP}.

\section*{Acknowledgement}
The author is a JSPS Fellow for Research Abroad.
He thanks Atsuo Kuniba and Kazumitsu Sakai
 for useful discussion.

\end{document}